\documentstyle[12pt,aaspp,psfig]{article}
\def\kms  {km s$^{-1}$}

\def\etal   {{\it et~al.}}
\def\nh {${\rm N_H}$}
\def\aox {$\alpha_{\rm ox}$}

\def\lax    {${_<\atop^{\sim}}$}
\def\gax    {${_>\atop^{\sim}}$}
\def\aox {$\alpha_{\rm ox}$}

\def\kms    {~km~s$^{-1}$}

\begin{document}

\title{The X-ray Warm Absorber in NGC3516}
\author{Smita Mathur$^1$}
\affil{smita@cfa.harvard.edu}
\author{Belinda J. Wilkes$^1$}
\affil{belinda@cfa.harvard.edu}
\author{Thomas Aldcroft$^1$}
\affil{taldcroft@cfa.harvard.edu}
\setcounter{footnote}{1}
\footnotetext{Harvard-Smithsonian Center for Astrophysics,
60 Garden St., Cambridge, MA 02138}

\received{}
\accepted{}
\lefthead{Mathur, Wilkes \& Aldcroft}
\righthead{Warm Absorber in NGC~3516}

\section*{Abstract}
The Seyfert 1 galaxy, NGC3516 has been the subject of many absorption
line studies at both ultraviolet and X-ray wavelengths. In the UV,
strong, broad, variable associated metal line absorption with velocity
width $\sim 2000$ \kms\ is thought to originate in gas with \nh\ \gax
$10^{19}$ cm$^{-2}$ lying between 0.01 and 9 pc from the central
active nucleus.  The {\it Ginga} X-ray data are consistent with
several possibilities: a warm absorber and a cold absorber combined
either with partial covering or an unusually strong reflection
spectrum.  We present ROSAT observations of NGC3516 which show a
strong detection of a warm absorber dominated by a blend of
OVII/OVIII edges at $\sim$ 0.8 keV with \nh\ $\sim 7\times 10^{21}$ cm$^{-2}$
and U: 8--12. We argue that NGC3516 contains an outflowing `XUV'
absorber showing the presence of X-ray absorption edges, that are
consistent with the presence
of broad absorption lines in the old IUE spectra and their
disappearance in the new UV observations. Our dynamical model
suggests that the OVII absorption edge will continue to
weaken compared to the OVIII edge, an easily testable
prediction with future missions like AXAF.
Eventually the source
would be transparent to the X-rays unless a new absorption system is
produced.

\newpage
\section{Introduction}
Given the recent success of efforts to unify associated ultraviolet (UV)
metal line  absorbers with ionized (``warm'') absorbers
seen in the X-ray (Mathur
\etal\ 1994, 1995, Mathur 1994), NGC3516 is a prime candidate for such
an investigation.
NGC~3516 is a low-redshift (z=0.0089) Seyfert 1 galaxy with a
variable, associated UV
absorption line system which has been extensively monitored with IUE
(Voit et
al 1987, Walter et al 1990, Kolman \etal\ 1993 (KEA); Koratkar \etal\ 1996).
Based on IUE spectra, NGC~3516 contains the strongest (largest
equivalent width)
UV absorption system known in a Seyfert 1 galaxy. This system contains
at least two distinct components: a
broad (FWHM$\sim$2000 \kms) variable component, most likely to be associated
with the X-ray absorber, and a narrow ($\sim$ 500 \kms)
non-variable
component.  The equivalent-width of the CIV $\lambda 1549$ broad+narrow
absorption line varies by a factor of 4-5  and anti-correlates with the UV
continuum level on short timescales of days.
 Walter \etal\ (1990) have also found variations on longer timescales
($\sim$ years) possibly due to the absorbing cloud passing through the
line of sight.
Recent observations have found
that the broad variable absorption lines of CIV and NV have
disappeared (Koratkar \etal~1996, Kriss \etal~1996b).

The X-ray absorber is not so well-known or heavily studied. The
highest quality data to date are from 1989 {\it Ginga} observations
reported by KEA, consistent with several possibilities: a warm
absorber, a cold absorber with partial covering, or a cold absorber
with an unusually strong reflection spectrum. If the absorption is due
to the cold matter, the column density is large N$_H \sim 10^{22}$
cm$^{-2}$.  These authors concluded that a common origin for the X-ray
and UV absorbing material is not possible due to the apparently
different ionization states and column densities of the UV and X-ray
absorbers. {\it Ginga} data were later re-analyzed by Nandra \& Pounds
(1994), showing that a warm absorber describes the data well.  New
ASCA observations of NGC3516 are presented in a recent preprint by
Kriss \etal~(1996a),
which show a presence of an ionized absorber similar to the one
discussed in this paper.

Here we present a long ($\sim$ 1 day) high signal-to-noise ROSAT
(Trumper 1983) Position Sensitive Proportional Counter (PSPC,
Pfefferman et al. 1987) spectrum of
NGC3516 obtained in 1992 which, with its lower energy range, is sensitive
to absorption edges due to ionized oxygen. The high S/N allows the
strong detection of both
OVII and OVIII edges independently, in spite of the limited spectral
resolution of the PSPC. These edges show the presence of an ionized
absorber in
X-rays. We
discuss the constraints these data place on the physical
conditions in the absorber and revisit the question of a common origin
for the X-ray and UV absorbing material.

\section{{\it ROSAT} Observations and Data Analysis}

 NGC 3516 was observed with the {\it ROSAT} PSPC on
October 1, 1992
for a total livetime of 13,081 sec. over a real time span of 83,455 sec.  We
retrieved the data from the HEASARC\footnote {High Energy Astrophysics
Science Archive Research Center is a service of the Laboratory for
High Energy Astrophysics (LHEA) at NASA/GSFC.} database and analyzed it
using the PROS\footnote {Post-Reduction Offline Software}
package in IRAF. The source counts were extracted
from within a 3$^{\prime}$  radius circle centered on the source
centroid. The small background was estimated from an annulus centered on the
source and inner and outer radii of 5$^{\prime}$ and 7$^{\prime}$
 respectively. The total net counts were  59,225$\pm$246 yielding
a count rate of 4.53$\pm$0.02 s$^{-1}$ (Table 1) (c.f. $1420\pm340$ counts for
3C351 in which absorption edges were detected, Fiore et al. 1993).

Since  NGC 3516 is known to be a variable X-ray source (Halpern 1982), we
looked for X-ray flux variations over the total span of the
observation. A total of eight good time intervals (GTI) were
found.
The net counts in each GTI were extracted in the same way as
discussed above. The live time, count rate and the total counts in
each GTI are given in Table 1. The change in count rate shows the flux
variability which is clearly seen in the light curve in Figure 1.
The variations are significant at the 99.99\% level ($\chi^2$ test,
K-S test using PROS task ``vartst'').
Since variability of $\sim$40\% is clearly seen, the source spectrum
for each GTI was extracted
separately, and analyzed using XSPEC\footnote {X-ray SPECtral
analysis software available from NASA-GSFC}.
As can be seen in Table 1, the individual GTIs contain
sufficient counts to fit complex spectra.

\subsection {Spectral Analysis}

 The results of the spectral analysis of all the datasets are
presented in Table 2. All the spectral fits were made to the 3--34 PHA
channels as extracted by the standard PROS analysis. Channels 1 and 2
were ignored since they are inadequately calibrated.  The
response matrix released in January 1993 was employed. The errors
represent the 90\% confidence interval.

 Our first fit used a simple power law with absorption by cold
material at zero redshift with solar abundances, and cross sections by
Morrison and McCammon (1983). Both the power-law slope and absorption
column density were free to vary.  In all GTIs, the column density of
the cold absorber (N$_H$) was consistent (within 90\% confidence) with
the Galactic value of $3\times 10^{20}$~ atoms~cm$^{-2}$ (Heiles 1975)
(e.g. N$_H=3.0\pm0.1\times 10^{20}$~cm$^{-2}$ in GTI 4; and
N$_H=2.4\pm0.5 \times 10^{20}$~cm$^{-2}$ in GTI 7). The single
power-law (and all subsequent) fits were redone with N$_H$ fixed at
the Galactic value and the results are presented in Table 2.

The single power-law fits are unacceptably bad, with $\chi^2_{\nu}$
\gax 4 and significant negative residuals between 0.7 and 1 keV (see
Figure 2). This is a clear signature of K-shell absorption edges due
to OVII and/or OVIII.  We then fitted the spectra with a power-law
(PL), cold Galactic absorption, and an edge at the energy of
a redshifted OVII K-edge (E(rest frame)=0.74 keV, z=0.0089). The
improvement in the fit was dramatic, as demonstrated by the results of
an F-test (Table 2), although small residuals remained near the edge
in several GTIs. For consistency we then fitted the spectrum of all
GTIs with an additional edge at the energy of the redshifted OVIII
K-edge (E(rest frame)=0.87 keV).  Once again the improvement in the
fit was significant and the spectrum was well fit ($\chi^2_{\nu} =
1.1$, Table 2, Figure 3) except for GTI 4 ($\chi^2_{\nu} = 1.6$, see
below).  Thus, even though the energy resolution of the ROSAT PSPC is
insufficient to resolve the OVII/OVIII edges, the absorption in NGC
3516 is so strong ($\tau \sim 1$) that the data {\it require} two
separate edges in half of the GTIs ($>99$\% confidence, labeled
``$^c$" in Table 2). These
are not simply the four GTIs with highest S/N.

 As discussed above, the data were consistent with
 N$_H$=N$_H$(Galactic).
 To confirm that this is the case even with the `power-law plus two
edges' model, we fit the spectra with a PL,
Galactic N$_H$ (fixed), two edges and an additional cold N$_H$ at the
source. The fitted value of the additional column density was always
much smaller than Galactic ({\it e.g.} 1.2$\times 10^{18}$
atoms~cm$^{-2}$ for GTI 1) and consistent with zero. Excess N$_H$ is
clearly not required.
GTI 4 was the only exception for which an excess absorption
N$_H=4.9\pm 2.6 \times 10^{19}$  improved the fit
significantly ($\chi^2_{\nu} = 1.28$, improvement $>$ 99\%
significant, F-test). Parameters for this best fit model are given in
the last line for GTI 4 in Table 2. Since GTI 4 has the maximum S/N, it is
possible that a small additional column is present (N$_H$\lax
$3.7\times 10^{19}$ ~cm$^{-2}$) but undetected in
the remaining GTIs.

 Since the flux is found to be variable but the absorption and
spectral shapes are not (see section 3.1), we also fitted all the GTIs
together with a `power-law plus two edges' model allowing their
normalizations to be free (Table 2). Excess absorption was not
required, though the best value was N$_H=3.3^{+1.3}_{-3.3} \times
10^{19}$ ~cm$^{-2}$, similar to the results above.

\section{The X-ray warm absorber}

 Knowing the total column density and the opacity of the oxygen edges,
we can constrain the ionization state of the warm absorber (Mathur,
Elvis \& Wilkes, 1995). The present case of NGC3516 is a little tricky,
however, because N$_H$ is not known.
Instead we need to estimate the N$_H$ and the ionization parameter (U) by
fitting the spectrum with a warm absorber model. Alternatively
we can derive these parameters in a more elegant way, exploiting the
fact that we observe {\it both} OVII and OVIII edges in NGC3516. We apply
both these methods below.

 Let $\tau_{OVII}$ and $\tau_{OVIII}$ be the opacities due to K-edges
of OVII and OVIII respectively. The absorption cross-sections of OVII
and OVIII are 0.28$ \times 10^{-18}$ cm$^{-2}$ and 0.098 $ \times
10^{-18}$ cm$^{-2}$ respectively (CLOUDY; Ferland 1991). Using these
we can derive the column densities of the two ions, N$_{OVII}$ and
N$_{OVIII}$ and the ratio N$_{OVII}$/N$_{OVIII}$ = f$_{OVII}$/f$_{OVIII}$
where f$_{OVII}$ and f$_{OVIII}$ represent the fraction of oxygen in these
two stages of ionization. Over the span of ROSAT observations (GTI 1
through 8), f$_{OVII}$/f$_{OVIII}$ ranges from $\sim 0.16$ to 5.8.
We determined the dependence of f$_{OVII}$ and f$_{OVIII}$
on the ionization parameter, U, for a photoionized gas cloud using CLOUDY  and
assuming solar abundances and a standard AGN continuum (Ferland 1991,
version 80.06). The density was assumed to be $10^5$~cm$^{-3}$.
The results are displayed in Figure 4 and demonstrate that the observed
range in f$_{OVII}$/f$_{OVIII}$ occupies a small
area in the parameter space of fractional ionization as a function of
ionization parameter.
Using Figure 4, we can not only infer
the ionization parameter (U ranges from $\sim 3.2$ at GTI 6 to $\sim
18$ at GTI 3), but also f$_{OVII}$ and
f$_{OVIII}$. These lie in the range 0.06$<f_{OVII}<$0.8, and
0.1$<f_{OVIII}<$0.5.  Using the abundance of oxygen relative to hydrogen
(8.51$\times 10^{-4}$; Grevesse \& Andres 1989) and f$_{OVII}$ or
f$_{OVIII}$, the total column density is readily calculated to be N$_H
= 0.4 - 2.2 \times10^{22}$ cm$^{-2}$.

As discussed above, the ionization parameter of the X-ray absorber is
determined using Figure 4. It should be noted however that the exact
relation of ionization fractions and U depends upon the shape of the
input continuum (Fiore \etal\ 1993, Mathur \etal\ 1994).  We have used
a standard AGN continuum (CLOUDY, Ferland 1991) in our analysis.  With
a different continuum shape, the inferred range of U would change but,
qualitatively, Figure 4 would remain the same. For example, an X-ray
slope of 1.0 (as seen in the ROSAT data) rather than the 0.7 assumed
in the standard continuum
leads to a larger value of U: 8.5\lax\ U
\lax\ 23.
On the other hand if the X-ray slope is much flatter ($\alpha \sim
0.3$) as seen by {\it Ginga}, U would be correspondingly lower.
Independently of the shape of the input continuum, the X-ray data for
the absorber in NGC3516 requires f$_{OVII}$/f$_{OVIII} \sim 1$,
corresponding to a highly ionized absorber.  A similar result was
obtained by Kriss \etal~(1996a) who used a variety of incident
continua and found that their results were independent of the exact
shape of the continuum apart from the deduced ionization parameter U.
The only derived quantity that depends upon the exact value of U (and
so on the shape of the continuum) is the distance of the absorber from
the central continuum source (R$^2=Q/{4\pi U n c}$). Since there is a
four orders of magnitude uncertainty in density, n (section 3.1), the
uncertainty in R is very large, far exceeding that due to the
uncertainty in U. We will thus assume the standard continuum in the
rest of the paper.  Kriss \etal\ (1996a)  use a continuum with significantly
flatter \aox\ and correspondingly lower value of U=1.66 in their
analysis of 1995 ASCA data,  so we cannot
directly compare our U value with Kriss \etal~(1996a).

Unlike the other objects showing an X-ray warm absorber (e.g. 3C351:
Fiore \etal~1993, NGC3783: Turner \etal~1994, NGC5548: Mathur
\etal~1995), the X-ray continuum in NGC3516 is transparent at low
energies.  To investigate this absence of a low energy turnover in the
ROSAT spectrum of NGC3516, we generated the transmitted spectrum for
the best fit U and N$_H$ (using CLOUDY). Figure 5 shows the input
spectrum (solid line; ``standard'' AGN spectrum) and the transmitted
spectrum (dotted line) over the ROSAT and {\it Ginga} energy bands.
It is clearly seen that the spectrum is transparent at low energies
and the OVII/OVIII absorption edges are prominent in the ROSAT energy
range.

 As a consistency check we also fitted the spectra with a warm
absorber model. The models were generated using CLOUDY and then
incorporated into XSPEC as `table models'. The input parameters to
CLOUDY were
same as used above. The results of the fits are
given in Table 3 and are consistent with the results discussed above.
When N$_H$ is fixed to the Galactic value, the fits are acceptable for
all the GTIs except GTI 4, similar to the results in section 2.1 (first line
for each GTI in Table 3).  When N$_H$ was allowed to be free, the fits
improved and were good in all the GTIs (second line in Table 3). The
best fit N$_H$ was slightly larger than but consistent with the
Galactic except in GTI 4, again similar to the results above. In these
models the range of the ionization parameter is $3<U<28$ and the
column density of the warm absorber is N$_H$=0.4--3.2$\times10^{22}$
cm$^{-2}$, consistent with the results of the empirical fits.

\subsection{Variability}

 As mentioned previously, the flux level of NGC3516 is variable as a
function of time (Figure 1). However, the variability of opacities in
OVII and OVIII absorption edges is not significant (Figure 1, \lax
2$\sigma$ variations).
 So all the GTIs were fitted together, leaving relative normalization
free, resulting in a tighter constraint on U (6\lax U\lax 13, 90\%
confidence). If the entire data set is fitted (i.e. without dividing
into separate GTIs), then the range in U is even smaller (7.9\lax
U\lax 12.6, 90\% confidence) and N$_H$=0.7$\pm0.1 \times10^{22}$
cm$^{-2}$ (Table 3). We use these values of U and N$_H$ in the rest of
the paper.

 The ionization parameter is expected to increase with increasing flux
on the photoionization time scale. The OVII photoionization time scale
is t$_{ph}= 2\times 10^5 n_6^{-1}$~ seconds (see e.g. Reynolds
\etal~1995). The lack of variability in OVII the opacity implies that
t$_{ph}$\gax 10$^4$ seconds. This puts an upper limit on the density of
the absorber,  n\lax $2\times 10^{7}$ cm$^{-3}$.

In the UV, variations of absorption
line strengths on timescales of days have been discussed for the CIV
absorber and used to derive a density of n\gax 10$^{5}$ cm$^{-3}$
in the absorbing region (Voit \etal\ 1987). This calculation, however,
assumes that CIV is the dominant ionization state. In the present
case, a correction factor of n$_{CIV}$/n$_{CV}$ needs to be applied to
the recombination time scale. The resulting lower limit on the density
of the absorber is then smaller, n\gax 10$^{3}$ cm$^{-3}$.

\section{Comparison with the {\it Ginga} results}

 The power-law slope in the ROSAT PSPC data is much steeper ($\alpha
\sim 1$) than that seen by {\it Ginga} ($0.25<\alpha<0.43$, KEA).
Since the PSPC range is dominated by the absorber, the {\it Ginga}
slope is more likely to indicate the true power-law slope.  The ROSAT
PSPC data show no significant excess cold absorption, but require a
warm absorber with \nh\ $\sim 7\times 10^{21}$ cm$^{-2}$ and 8\lax
U\lax 13. It can be seen from Figure 5 that this model does {\it not}
predict the presence of an Fe edge ($\tau_{pred}$\lax 0.025). This is
inconsistent with the {\it Ginga} data which showed a ``cold''
absorber (\nh\ $\sim 10^{22}$ cm$^{-2}$) and an Fe edge. As can be
seen in Figure 5, the ionized absorber causes a low energy turn-over
within the {\it Ginga} band which would mimic ``cold absorption'' in
the {\it Ginga} data. Indeed, re-analysis of the {\it Ginga} data
showed that the low energy turn-over is most likely due to the warm
component (Nandra \& Pounds 1994).
However, the origin of the Fe-K edge would have to be different from
the XUV absorber discussed here. It may be,
e.g.,  from the torus (Krolik and Kriss 1995) which possibly
grazes our line of sight. Alternatively, the Fe edge and the large
 column may be associated with an absorption system that has moved out
of the line of sight.
We note in passing that there is no obvious Fe edge in recent ASCA
observations (Kriss \etal\ 1996a).

\section{Comparison of X-ray and UV absorbers}

 In many AGN, the X-ray  and UV absorbers have been found to be one and the
same (the `XUV Absorbers': Mathur \etal~1994, 1995). In these cases
apparent inconsistencies between the conditions in the UV and X-ray
were resolved by realizing that the ions observed in the UV were not
the dominant ions in the gas.  NGC 3516 also
shows both X-ray and UV absorbers. In a previous
study based on the simultaneous (October 1989) {\it Ginga} and IUE
observations, KEA
concluded that, if an X-ray ionized absorber is present, it is unlikely to be
due to the same gas as the UV absorber. The
combination of the strength of the UV CIV absorption and the very high
ionization state of Fe (\gax 12, consistent with $>5 \times$ ionized C)
indicated by the ionized absorber model fit to the {\it Ginga} data
implied to the authors that a consistent solution was unlikely, although
no detailed calculations were presented. Since the ionized absorber
observed with ROSAT is clearly different from the {\it Ginga}
absorber, the question of an XUV absorber in NGC3516 should be
revisited.

 Here we investigate quantitatively whether the ROSAT ionized absorber
is consistent with the broad UV absorber with high ionization lines.
We use the method described in Mathur \etal~(1995). The ROSAT
observations were made in 1992. There were no simultaneous UV
observations. The 1993 IUE observations  show that the broad UV
high ionization absorption lines had disappeared (Koratkar
\etal~1996). Figure 4 shows the model with fractional ionization of
CIV as a function of U. If the
ionization parameter of the X-ray absorber were toward the high end of the
observed range (U=12.8), the ionization fractions of CIV and NV would be
very small ($\log$ f \lax$-4$). Assuming solar abundances of carbon and
nitrogen
(3.63$\times 10^{-4}$ and $1.12\times10^{-4}$ respectively, Grevesse \& Andres
1989) and the total column density of $7\times 10^{21}$ cm$^{-2}$ as
derived from the X-ray data gives  N(CIV) $< 10^{14}$
cm$^{-2}$ and N(NV) $< 8\times10^{13}$ cm$^{-2}$. The absorption
lines would not be detected. In this case there would be an X-ray
ionized absorber, but no broad high ionization UV lines of CIV or NV,
consistent with the observations. However, the X-ray absorber must
produce a detectable OVI $\lambda$1034 line with $-2.1$\lax $\log$
f$_{OVI}$ \lax $-2.7$ (Figure 4). Since there were no contemporaneous far-UV
observations with  ROSAT, this cannot be directly tested.
Analysis of absorption lines embedded in the emission line profiles of
quasars is extremely difficult since the shape of the emission lines is
unknown. However, visual inspection of the 1995 HUT data for
NGC3516 (Kriss etal 1996b) shows a broad absorption
feature blue-wards of the OVI emission line peak and at roughly the expected
redshift of the broad UV absorber.

 It is also of interest to point out the position of the broad CIV and NV
absorption lines on the f-U curve when they were strong.  The
equivalent width of the variable broad CIV absorption line in the IUE
observations ranged from $\sim$3\AA~ to $\sim$10\AA ~(KEA) leading to
a lower limit on the CIV column density N$_{CIV} > 2.2\times 10^{15}$
cm$^{-2}$. Similarly Voit \etal~(1987) report N$_{CIV}$ \gax $10^{15}$
cm$^{-2}$. Assuming this lower limit gives a lower limit on the
ionization fraction of CIV: f$_{CIV}$\gax $4\times 10^{-4}$ ($\log$~
 f$_{CIV}$\gax -3.4). The NV EW given by KEA varies in the range 1.1
-- 7.1 \AA. This corresponds to N(NV) $> 3.5\times10^{14}$ cm$^{-2}$
(using oscillator strength f=0.235, Allen 1973). The lower limit on
the ionization fraction of NV is estimated to be f$_{NV}$\gax
$5\times 10^{-4}$ ($\log $ f$_{NV}$\gax -3.3).  The arrows in Figure
4 indicate these lower limits on f$_{CIV}$ and f$_{NV}$ observed by IUE.
These lie in  a range of U smaller than  that corresponding to the  ROSAT
observations. This suggests that if U increases as a function of time, both
the former presence and current absence of the broad UV lines would be
consistent with the observed X-ray warm absorber.

The IUE data have occasionally shown the presence of a noisy, possibly
broad SiIV absorption feature (Voit, Shull \& Begelman 1987).  The
conditions of the warm absorber derived here do not predict detectable
SiIV absorption ($\log$ f$_{SiIV}$\lax$-30$). However, SiIV absorption
is also present, with no significant change in strength, in the 1993
IUE spectra in which no broad, high ionization lines are reported
(Koratkar \etal~1996).  Thus it appears that the feature is a part of
the narrow UV absorption system rather than the XUV absorber.
Note that the two other objects with well-studied XUV absorbers
(e.g.3C351, NGC5548) do not have any such additional narrow absorption
systems.  It is possible that the second, low column density absorber
observed in the ROSAT data is related to this narrow UV system.

\section{Discussion}

 We argue that NGC3516 contains an XUV absorber. It has high column density
(N$_H=7\pm 1 \times 10^{21}$ cm$^{-2}$) derived from the X-ray
observations. The UV observations imply that the absorber is outflowing with a
velocity of $\sim 500$ km~s$^{-1}$ (given its blueshift) and the line
width implies an  internal
velocity dispersion of $< 2000$ km~s$^{-1}$ (see sec. 1). Assuming
the upper limit on its ionization parameter (U$<$13), a density of
$2\times 10^{5}$ cm$^{-3}$, and N$_H=7\times 10^{21}$ cm$^{-2}$, the
distance of
the absorber from the central continuum source is $< 5\times
10^{18}$~cm and its thickness is about $3.5\times10^{15}$~cm. At this high
value of U no detectable broad UV lines are predicted (sec. 3, fig.
4), consistent with the
  disappearance of the broad UV absorption lines ( Koratkar
\etal~1996, Kriss \etal~1996b).

 Kriss \etal (1996b) have presented higher resolution ASCA data of the
ionized absorber in NGC3516. Based on the non-detection of X-ray
absorption lines in the ASCA data they have argued that the velocity
dispersion parameter `b' of the absorber is not larger than $\sim 200$
km s$^{-1}$. This result is based on the models of Krolik \& Kriss
(1995) in which
resonance line scattering is an important process.  In this case the
X-ray absorber is narrow while the UV absorber is broad, implying that
the two cannot originate in the same gas.
We note however that their models covered the b range from 10 km s$^{-1}$
to 200 km s$^{-1}$, and did not definitively rule out the presence
of broad ($\sim 2000$ kms$^{-1}$)
blended features which could mimic a different continuum shape.  More
importantly, in similar models presented by Netzer (1996), absorption
and emission due to resonance line scattering almost identically
cancel out so that no absorption lines are expected even for
turbulances significantly larger than the thermal line widths. Thus,
the non-detection of absorption lines in ASCA data may simply be because
there are no lines there.  Clearly the theoretical situation is too
uncertain for the absence of features to provide constraints on the b
parameter of the absorbing gas.  Since the UV absorber has a measured
width of $\sim 2000$ km s$^{-1}$, we will adopt this value in our
ensuing discussion of a combined X-ray/UV absorbing region.

 We have seen that the X-ray absorber is consistent with both the
presence of the broad UV absorption lines in the past and their
subsequent absence, but can we develop a consistent physical picture
to explain this variation? The evolution of ionization parameter as
seen in figure 4 is suggestive.  The earliest known observations of the
broad UV absorption lines are from 1978. Strong lines were also
observed in 1989 (KEA), while they have disappeared since $\sim$ 1993.
Thus the large scale variability time scale of the absorber seems to
be some 3--6
years.  Given its outflow velocity, the distance traveled by the
absorber away from the ionizing continuum source in this time is
$5\times 10^{15}$~cm. Clearly, this is a small fraction of the
distance of the absorber from the central continuum and there would be
no significant change in ionization parameter corresponding to this
change in flux.  On the other hand, the velocity dispersion of the
absorber is large (FWHM$\sim 2000$ km s$^{-1}$).  If the absorber is
expanding at this rate, its thickness will change by $\sim 2-4 \times
10^{16}$~cm  in 3--6 years.
Thus, to have a present thickness of $\sim 3.5\times 10^{16}$~cm, the thickness
of the absorber 3 years ago must have been \lax$1.5\times 10^{16}$~cm
with a corresponding density of
\gax$5\times 10^5$~cm$^{-3}$. Then the ionization parameter in the
past could be as low as U=5,
 (see fig. 4) where the strength of the broad UV absorption lines
would be in the observed IUE range.
  CIV and NV lines would then be strong.
In this scenario the disappearance of the broad UV absorption lines is
due to increase in U caused by the drop in the density of the
absorber as it expands while outflowing from the ionizing source. For
this scenario to work, the fractional change in the thickness of the
absorber in 3--6 years should be significant. If the absorber is
dispersed in velocity space, the `b' parameter of each cloud would be
smaller. If we consider  a conservative estimate of $\sim
200$ km s$^{-1}$ of the velocity dispersion, a significant change in
the thickness of the absorber would   require a density in the
higher end of the allowed range (n\gax 10$^{6}$ cm$^{-3}$).

 The XUV absorber picture is thus completely consistent with the
observations: the presence of X-ray absorption edges, the presence of
broad absorption lines in the old IUE spectra and their disappearance
in the the new UV observations. In this scenario the broad UV
absorption lines will not reappear unless a new, similar absorption
system is generated. Over the next $\sim 15$ years the current
absorber will become more highly
ionized, the OVI absorption lines will disappear, and the  OVIII
edge will become stronger than the OVII edge.
Eventually, it will be completely transparent in X-rays and the
OVII/OVIII edges will  disappear. This is an easily testable
prediction with future missions. These estimates of time scale and
the amount of change in U are based on assuming b=2000 km s$^{-1}$ and
density of $\sim 2\times 10^5$ cm$^{-3}$. If the
internal velocity dispersion is smaller, the evolution described above
would be slower. We note that this  picture concerns long term
variations, we expect in addition  small scale variations due to flux
variability.

 As noted earlier, new simultaneous HUT
(UV) and ASCA (X-ray) observations are presented by Kriss
\etal~(1996a,b).
These data show an X-ray ionized absorber similar to that reported
here and narrow, associated absorption lines in the UV.
The authors are unable to reconcile the X-ray and UV absorbers
with a single absorbing region. This is entirely consistent with our
model predictions in which the X-ray warm absorber is instead
associated with the now invisible broad UV absorption. The absorption
systems  in NGC3516 are  clearly complex and multiple. The Kriss
\etal~(1996a,b) study shows
that each of the  X-ray and narrow UV absorbers require at least two separate
systems to match the data. More, simultaneous X-ray and UV data
allowing monitoring of the various components is necessary to
disentangle them. We cannot rule out the possibility that the X-ray
and UV absorbers are different. In that case the disappearance of the
broad absorption lines may be due to the clouds moving out of line
of sight as suggested by Walter \etal~(1990). Our XUV absorption
scenario offers an alternative explanation.

 An edge-on  orientation of 3C351, 3C212 and NGC5548 was inferred from
their lobe dominated radio structure. NGC3516 also has a steep radio
spectrum (Ulvastad and Wilson 1989) typical of lobe dominated AGN. Its
radio maps reveal an elongated, one sided, curved structure (Miyaji
\etal~1992). The geometrical structure of NGC3516 was inferred by Goad
and Gallagher (1987) by analyzing the velocity field of the
circumnuclear emission-line region. They found a bipolar outflow from
the nucleus in the plane of the sky so that  our line of sight grazes
the putative torus. Thus to date all active galactic nuclei shown to contain
XUV absorbers are edge on, consistent with the picture of outflowing
absorbers suggested by Glen, Schmidt, \& Foltz (1994).

\section{Conclusions}
ROSAT observations of NGC3516 in 1992 detected an X-ray warm absorber with
\nh\ $\sim 7\times 10^{21}$ cm$^{-2}$ and U $\sim$ 7.9--12.6.
Combining X-ray and UV observations imply that it is outflowing with a
velocity of $\sim 500$ km~s$^{-1}$  and has an internal
velocity dispersion of \lax 2000 km~s$^{-1}$. The constraints on
density are not very tight: $10^3$\lax n\lax $10^7$~cm$^{-3}$.
The distance of
the absorber from the central continuum source is \lax $5\times
10^{18}$~cm and its thickness is about $3.5\times 10^{15}$~cm assuming
density of $2\times 10^5$~cm$^{-3}$.
 The physical characteristics of the
absorber are consistent with all the observations: the presence of
X-ray absorption edges, the presence of broad
absorption lines in the old IUE spectra and their disappearance in the
the new UV observations. In this scenario of the XUV absorber in
NGC3516, the
 disappearance of the broad UV absorption lines is due to the
current high value of U caused by the drop in the density of the
absorber as it expands and moves away from the ionizing source. If
this dynamical model is correct, we expect the OVII absorption edge in
the X-rays to weaken compared to the OVIII edge. Eventually the source
would be transparent to the X-rays  as the
absorber continues to expand, unless a new absorption system is produced.

\begin*{Acknowledgements:~}
 It is our pleasure to thank our colleagues Martin Elvis and Paul
O'Brien for useful discussions. The financial support of the following
NASA contracts and grants is gratefully acknowledged: NAGW-4490
(LTSA), NAS5-30934 (RSDC), NAS8-39073 (ASC).
\end*

\newpage

\noindent
{\bf Figure 1:} NGC 3561 light curve (top panel), ~$\tau_{OVII}$
(middle panel),  ~$\tau_{OVIII}$ (bottom panel),  for the eight GTIs.
 \\

\noindent
{\bf Figure 2:} Residuals to the simple power-law fit to the entire data.
 The strong negative residuals between
0.7 and 1 keV are clearly visible. Channels 13--25 corresponding to
this energy range were ignored in the PL fit to show the absorption
edges clearly.\\

\noindent
{\bf Figure 3:} Data and the best fit spectrum (PL+2 Edges,
N$_H$ fixed to Galactic). \\

\noindent
{\bf Figure 4:} Ionization fractions f of OVI, OVII, OVIII, CIV and NV
as a function
of ionization parameter, U. The vertical lines define the
range of U for which the ratio f$_{OVII}$/f$_{OVIII}$ lies within the
observed range. The arrows on the CIV and NV curves
indicate the lower limits of f$_{CIV}$\gax $3\times 10^{-4}$
and f$_{NV}$ \gax $3.1\times 10^{-4}$ based on the published IUE data
(see text). \\

\noindent
{\bf Figure 5:} The input spectrum (solid line) and the transmitted
spectrum (dotted line) for N$_H=10^{22}$ cm$^{-2}$ and U=10. The
transmitted spectrum shows strong OVII/OVIII edges, but no Fe-K edge
or low energy cutoff. \\

\newpage
\begin{table}[h]
\caption{ROSAT Observations of NGC 3516}
\begin{tabular}{|lrrr|}
\hline\hline
Observation &  Net Counts & Exposure  & Net Count Rate \\
 &  & (s) & (s$^{-1}$) \\
\hline
Total & 59225$\pm$246 & 13081 & 4.53$\pm$0.02 \\
GTI 1 & 12860$\pm$127 &  3176 & 4.05$\pm$0.04 \\
GTI 2 &  4186$\pm$69  &  1150 & 3.64$\pm$0.06 \\
GTI 3 &  4068$\pm$71  &  1014 & 4.01$\pm$0.07 \\
GTI 4 & 14848$\pm$116 &  2900 & 5.12$\pm$0.04 \\
GTI 5 &  7007$\pm$96  &  1375 & 5.10$\pm$0.07 \\
GTI 6 &  7670$\pm$97  &  1612 & 4.76$\pm$0.06 \\
GTI 7 &  1027$\pm$39  &   232 & 4.43$\pm$0.17 \\
GTI 8 &  7571$\pm$97  &  1622 & 4.67$\pm$0.06 \\
\hline
\end{tabular}
\end{table}

\newpage
\begin{table}[h]
\caption{Spectral fits to ROSAT data of NGC 3516}
\begin{tabular}{|rlcccccc|}
\hline
Data & Model& $\alpha_E$ & Normalization$^a$ & $\tau_{OVII}$ &
$\tau_{OVIII}$ & $\chi^2$ (dof) & F$^b$ \\
&&&&&&&\\
\hline\hline
GTI 1$^c$& PL& 1.22$\pm 0.01$& 1.36$\pm 0.01$ &&& 132.2 (30)& \\
& PL+ 1Edge& 1.07$\pm 0.03$& 1.70$\pm 0.07$& 0.8$\pm 0.15$&& 43.55
(29)& 59.0 \\
& PL+ 2Edges& 1.04$\pm 0.035$& 1.76$\pm 0.07$ & 0.46$^{+0.19}_{-0.17}$&
0.45$\pm 0.19$& 31.11 (28) & 11.2\\
&&&&&&&\\
\hline
 GTI 2& PL& 1.18$\pm 0.04$& 1.24$\pm 0.04$& & & 56.5 (30)& \\
& PL+ 1Edge& 1.06$\pm 0.06$& 1.51$\pm 0.11$& 0.71$^{+0.27}_{-0.26}$& &
33.68 (29)& 19.6 \\
&  PL+ 2Edges& 1.05$\pm 0.06$& 1.53$\pm 0.12$& 0.62$\pm 0.35$&
0.11$^{+0.36}_{-0.11}$& 33.45 (28)& 0.2 \\
&&&&&&&\\
\hline
GTI 3$^c$& PL& 1.23$\pm 0.03$& 1.33$\pm 0.03$& & & 62.87 (30)& \\
& PL+ 1Edge& 1.08$\pm 0.06$& 1.69$\pm 0.13$& 0.9$\pm 0.3$& & 32.17
(29)& 27.7 \\
& PL+ 2Edges& 1.02$\pm 0.07$& 1.8$\pm 0.14$& 0.34$\pm 0.33$&
0.75$^{+0.35}_{-0.37}$& 22.00 (28) & 12.9 \\
&&&&&&&\\
\hline
GTI 4$^c$& PL& 1.26$\pm 0.01$& 1.69$\pm 0.02$& & & 154.1 (30)& \\
& PL+ 1Edge& 1.12$\pm 0.02$& 2.08$\pm 0.05$& 0.77$\pm 0.08$& & 59.57
(29)& 46.0 \\
& PL+ 2Edges& 1.09$\pm 0.03$& 2.16$\pm 0.09$& 0.43$^{+0.18}_{-0.16}$&
0.45$\pm 0.18$& 44.87 (28)& 9.2 \\
&Best
Fit$^d$&1.24$\pm$0.09&2.23$\pm$0.09&0.75$^{+0.29}_{-0.25}$&0.19$^{+0.24}_{-0.19}$&34.55
(27)& 8.1\\
&&&&&&&\\
\hline
 GTI 5& PL& 1.22$\pm 0.02$& 1.72$\pm 0.08$& & & 66.20 (30)& \\
& PL+ 1Edge& 1.09$\pm 0.04$ &2.09$\pm 0.11$& 0.72$^{+0.20}_{-0.19}$& &
25.16 (29)& 47.3 \\
& PL+ 2Edges& 1.08$\pm 0.05$& 2.13$\pm 0.13$& 0.57$^{+0.30}_{-0.25}$&
0.20$^{+0.28}_{-0.20}$& 23.91 (28)& 1.5 \\
&&&&&&&\\
\hline
 GTI 6& PL& 1.16$\pm 0.02$& 1.55$\pm 0.03$& & & 117.4 (30)& \\
& PL+ 1Edge& 1.07$\pm 0.04$& 2.1$\pm 0.1$& 1.1$\pm 0.2$& & 20.89 (29)&
134.0 \\
& PL+ 2Edges& 1.06$\pm 0.04$& 2.1$\pm 0.1$& 1.0$\pm 0.3$&
0.06$^{+0.3}_{-0.06}$& 20.81 (28)& 0.1 \\
&&&&&&&\\
\hline
\end{tabular}
\end{table}

\newpage
\begin{table}[h]
Table~2:~Continued. \\
\begin{tabular}{|llcccccc |}
\hline
 GTI 7& PL& 1.2$\pm 0.1$& 1.5$\pm 0.1$& & & 19.91 (30)& \\
& PL+ 1Edge& 1.0$\pm 0.13$& 2.0$\pm 0.3$& 1.0$^{+0.7}_{-0.6}$& & 11.83
(29)& 19.8 \\
& PL+ 2Edges& 1.0$\pm 0.1$& 1.99$\pm 0.35$& 0.73$^{+0.36}_{-0.68}$&
0.3$^{+0.8}_{-0.3}$& 11.5 (28)& 0.8 \\
&&&&&&&\\
\hline
GTI 8$^c$& PL& 1.19$\pm 0.02$& 1.60$\pm 0.03$& & & 71.16 (30)& \\
& PL+ 1Edge& 1.06$\pm 0.04$& 1.9$\pm 0.1$& 0.7$\pm 0.2$& & 29.08 (29)&
42.0 \\
& PL+ 2Edges& 1.03$\pm.0.05$& 2.01$\pm 0.11$& 0.4$^{+0.25}_{-0.22}$&
0.4$^{+0.26}_{-0.27}$& 22.74 (28)& 7.8 \\
&&&&&&&\\
\hline
 GTI 1-8 &PL+ 2Edges& 1.06$\pm0.02 $& & 0.53$\pm 0.09$&
0.37$\pm0.09$& 244.6 (245)&  \\
         &PL+ 2Edges& 1.16$\pm 0.04$& & 0.74$^{+0.15}_{-0.13} $&
0.19$^{+0.12}_{-0.13}$& 226.9 (244)&  \\
& (N$_H$ free) &&&&&&\\
&&&&&&&\\
\hline
\end{tabular}
\smallskip

a: in units of $10^{-2}$ photons keV$^{-1}$ cm$^{-2}$ s$^{-1}$ at
1 keV\\
b: Parameter of the F-test. Note that  F \gax 7.5 is 99\%
significant and F \gax 13 is 99.9\% significant \\
c: datasets for which two edges are required by the data \\
d: PL+ 2Edges+ Excess absorption at the source (see text).
\end{table}

\newpage

\newpage
\begin{table}[h]
\caption{Warm absorber model fits to ROSAT data of NGC 3516}
\begin{tabular}{|rccccc|}
\hline
Data & $\alpha_E$ & Normalization$^a$ & N$_H$$^b$ (warm) &
 $\log U$ & $\chi^2$ (dof)  \\
&&&&&\\
\hline\hline
 GTI 1& 0.91$^{+0.06}_{-0.08}$& 2.15$^{+0.23}_{-0.17}$&
1.26$^{+0.34}_{-0.42}$& 1.34$^{+0.07}_{-0.12}$& 37.02 (28)\\
      & 1.1$^{+0.2}_{-0.1}$ & 2.22$^{+0.15}_{-0.17}$ & 0.77$^{+0.42}_{-0.28}$ &
1.09$^{+0.2}_{-0.25}$ & 30.41 (27)\\
&&&&&\\
 GTI 2& 0.95$\pm{0.11}$ & 1.79$^{+0.28}_{-0.22}$ & 0.69$\pm 0.37$ &
1.16$^{+0.23}_{-0.44}$ & 32.56 (28) \\
      & 1.4$\pm 0.3$& 2.2$^{+0.6}_{-0.4}$& 0.4$^{+0.2}_{-0.1}$&
0.45$^{+0.57}_{-0.45}$& 26.41 (27)\\
&&&&&\\
 GTI 3& 0.82$\pm 0.65$ & 2.37$^{+0.48}_{-0.03}$ & 3.19$\pm 0.85$ &
1.49$^{+0.01}_{-0.03}$ & 23.80 (28)\\
      & 0.82$^{+0.3}_{-0.2}$& 2.41$^{+0.5}_{-0.2}$& 3.35$^{+0.79}_{-3.34}$&
1.49$^{+0.008}_{-0.65}$& 23.52 (27)\\
&&&&&\\
 GTI 4& 0.9$^{+0.5}_{-0.1}$ & 2.76$\pm 0.12$ & 2.1$^{+1.2}_{-2.05}$ &
1.46$^{+0.03}_{-0.04}$ & 50.72 (28) \\
      & 1.3$^{+0.3}_{-0.1}$& 2.9$^{+0.5}_{-0.2}$& 0.6$^{+0.3}_{-0.2}$&
0.9$^{+0.2}_{-0.5}$& 31.54 (27)\\
&&&&&\\
 GTI 5& 0.94$\pm 0.56$ & 2.59$^{+0.58}_{-2.56}$ & 1.2$^{+1.4}_{-1.2}$
& 1.34$^{+0.15}_{-0.20}$ & 26.94 (28) \\
      & 1.3$^{+0.5}_{-0.2}$& 2.81$^{+0.9}_{-0.3}$& 0.5$^{+0.4}_{-0.1}$&
0.81$^{+0.3}_{-0.8}$& 17.35 (27)\\
&&&&&\\
 GTI 6& 0.95$\pm 0.48$ & 2.54$\pm{0.09}$ & 0.90$^{+0.5}_{-0.89}$ &
1.14$^{+0.19}_{-0.18}$ & 32.00 (28)\\
      & 1.3$^{+0.7}_{-0.2}$& 2.9$^{+1.6}_{-0.3}$& 0.6$^{+0.3}_{-0.2}$&
0.8$^{+0.3}_{-0.8}$& 21.45 (27)\\
&&&&&\\
 GTI 7& 0.87$^{+0.36}_{-0.25}$ & 2.44$^{+0.99}_{-0.64}$ &
1.03$^{+2.24}_{-1.0}$ & 1.22$^{+0.27}_{-1.22}$ & 11.98 (28)\\
      & 1.0$^{+1.1}_{-0.7}$& 2.5$^{+2.5}_{-0.7}$& 0.8$^{+2.9}_{-0.7}$&
1.06$^{+0.4}_{-1.06}$& 11.77 (27)\\
&&&&&\\
 GTI 8& 0.86$\pm 0.49$ & 2.57$\pm 0.43$ & 1.93$^{+1.71}_{-1.92}$ &
1.45$^{+0.04}_{-0.1}$ & 20.71 (28) \\
      & 1.1$^{+0.2}_{-0.3}$& 2.6$^{+0.2}_{-0.3}$& 0.8$^{+1.6}_{-0.4}$&
1.1$^{+0.3}_{-0.4}$& 19.05 (27)\\
&&&&&\\
 Total& 1.21$\pm 0.06$& 2.54$^{+0.1}_{-0.08}$ & 0.7$\pm0.1$&  1.0$\pm0.1$&
28.01 (27)\\
&&&&&\\
\hline
\end{tabular}
\smallskip

a: in units of $10^{-2}$ photons keV$^{-1}$ cm$^{-2}$ s$^{-1}$ at
1 keV\\
b: in units of 10$^{22}$ atoms cm$^{-2}$
\end{table}

\newpage
\begin{figure}
\centerline{
\psfig{figure=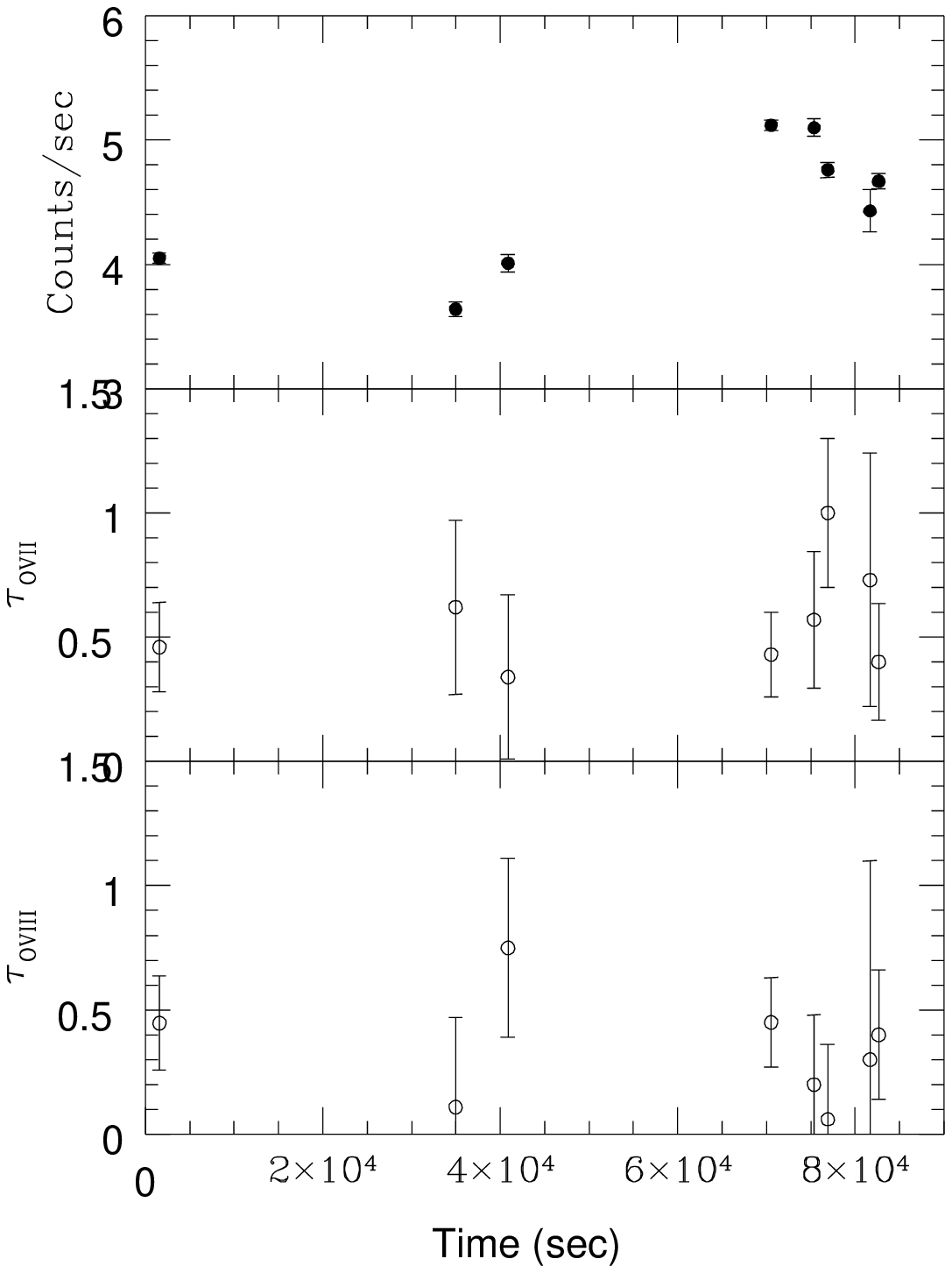,height=6.0truein,width=6.0truein}
}
\caption{ }
\end{figure}

\newpage
\begin{figure}
\centerline{
\psfig{figure=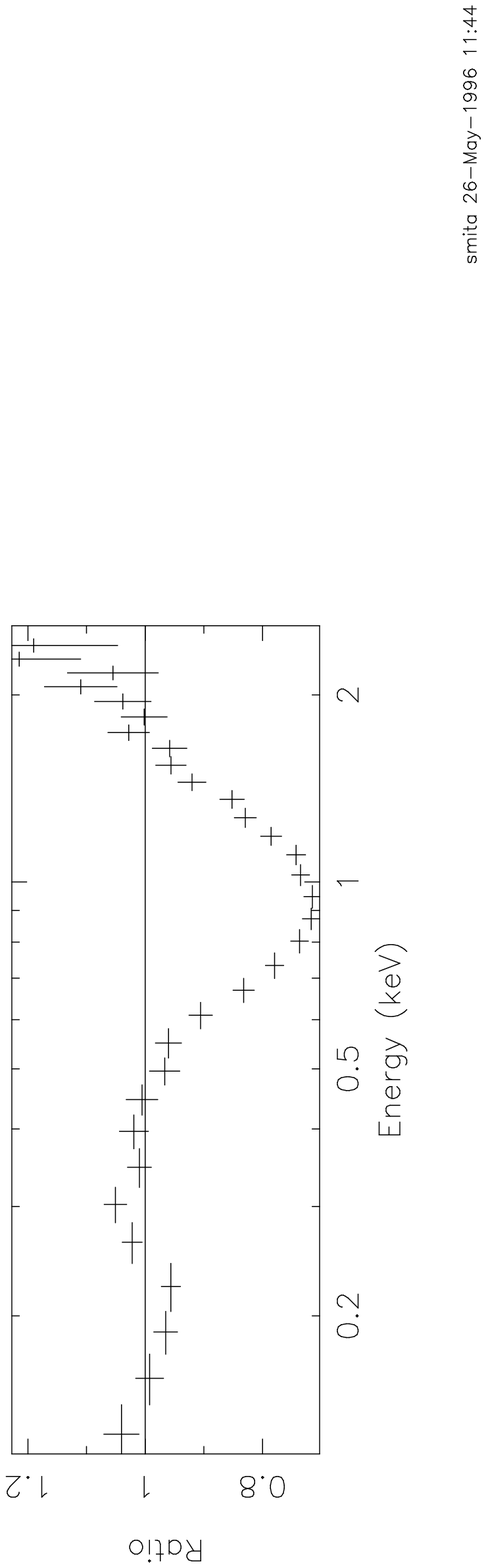,height=7.5truein,width=6.5truein}
}
\caption{ }
\end{figure}

\newpage
\begin{figure}
\centerline{
\psfig{figure=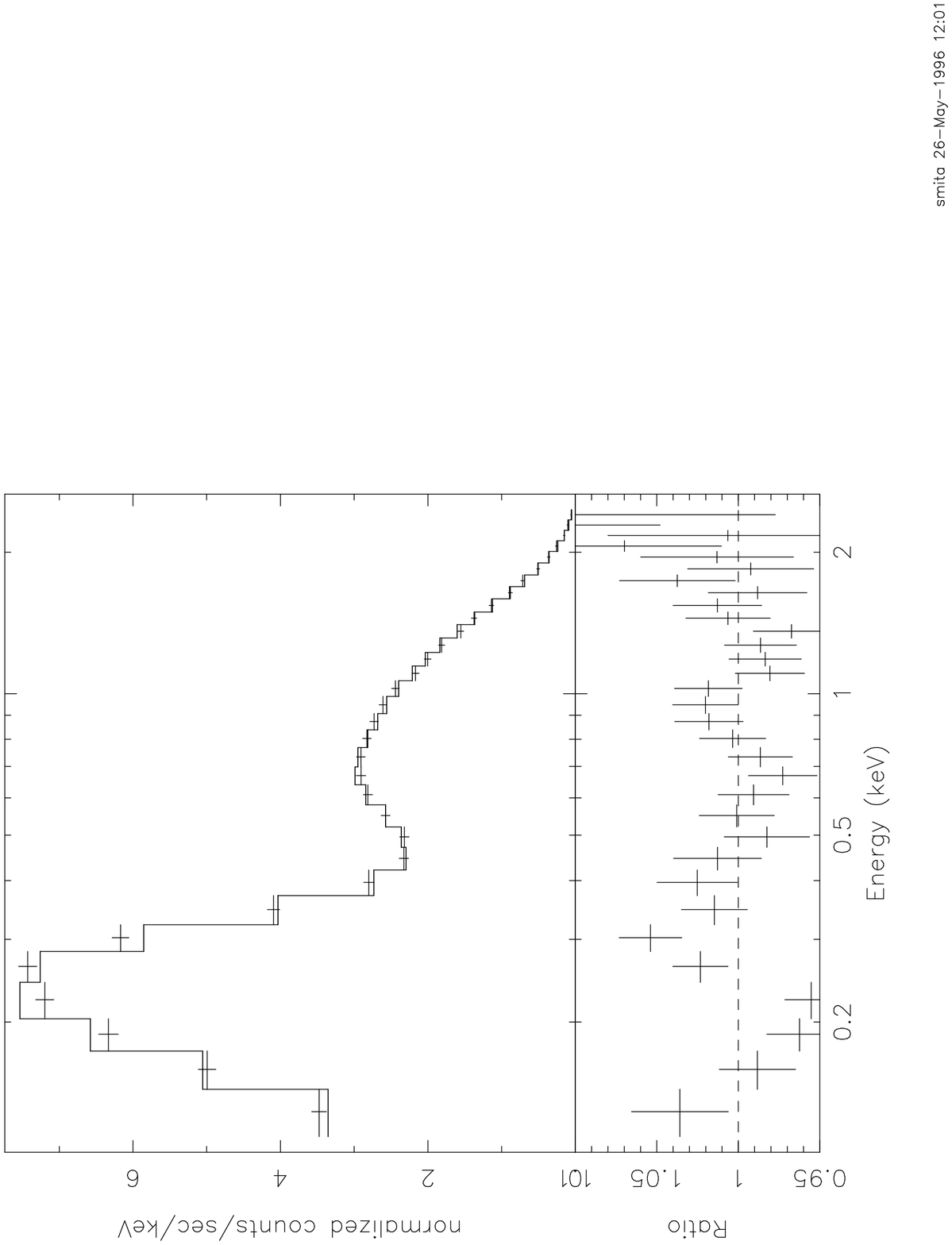,height=7.5truein,width=6.5truein}
}
\caption{ }
\end{figure}

\newpage
\begin{figure}
\centerline{
\psfig{figure=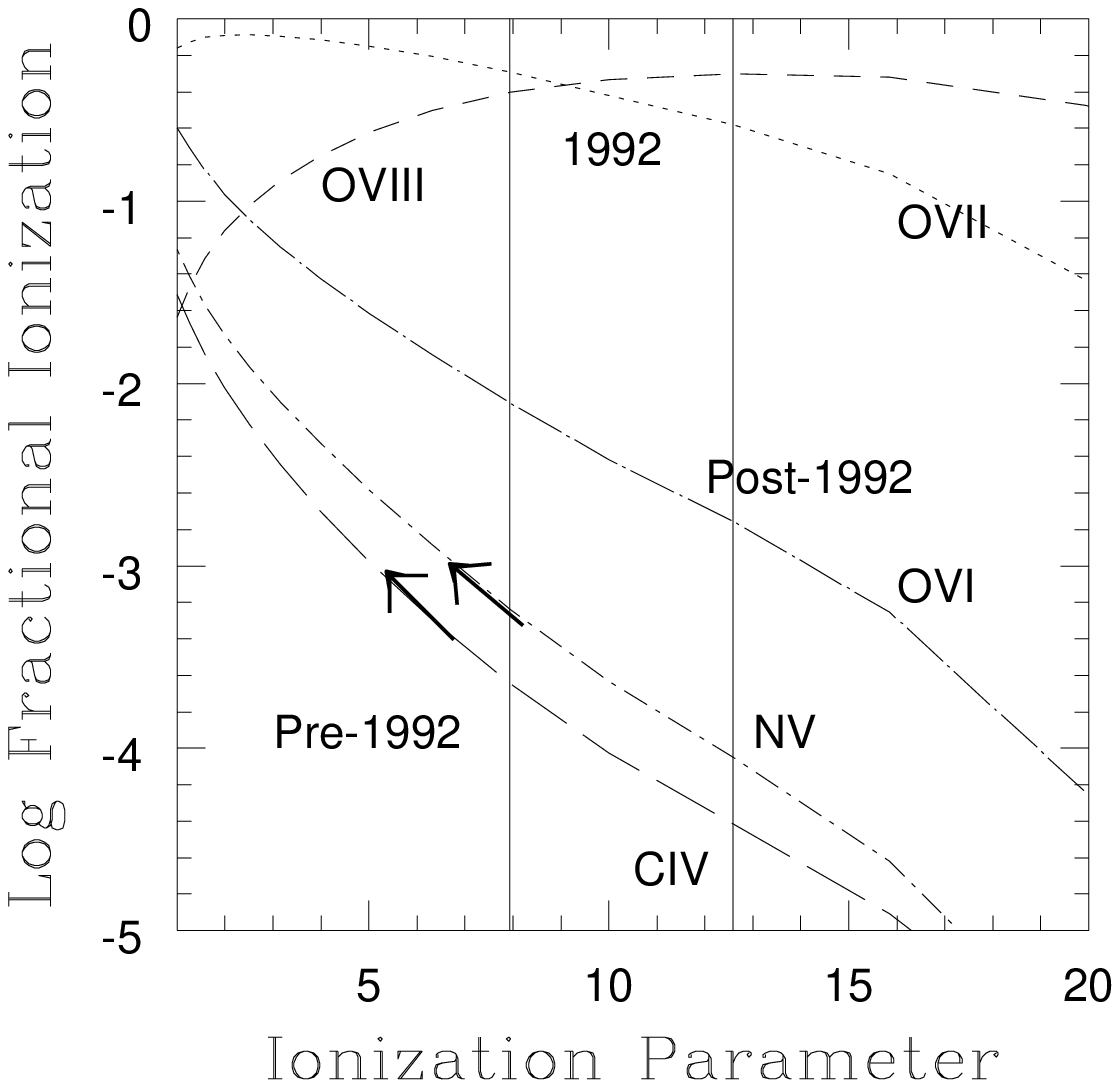}
}
\caption{ }
\end{figure}

\newpage
\begin{figure}
\centerline{
\psfig{figure=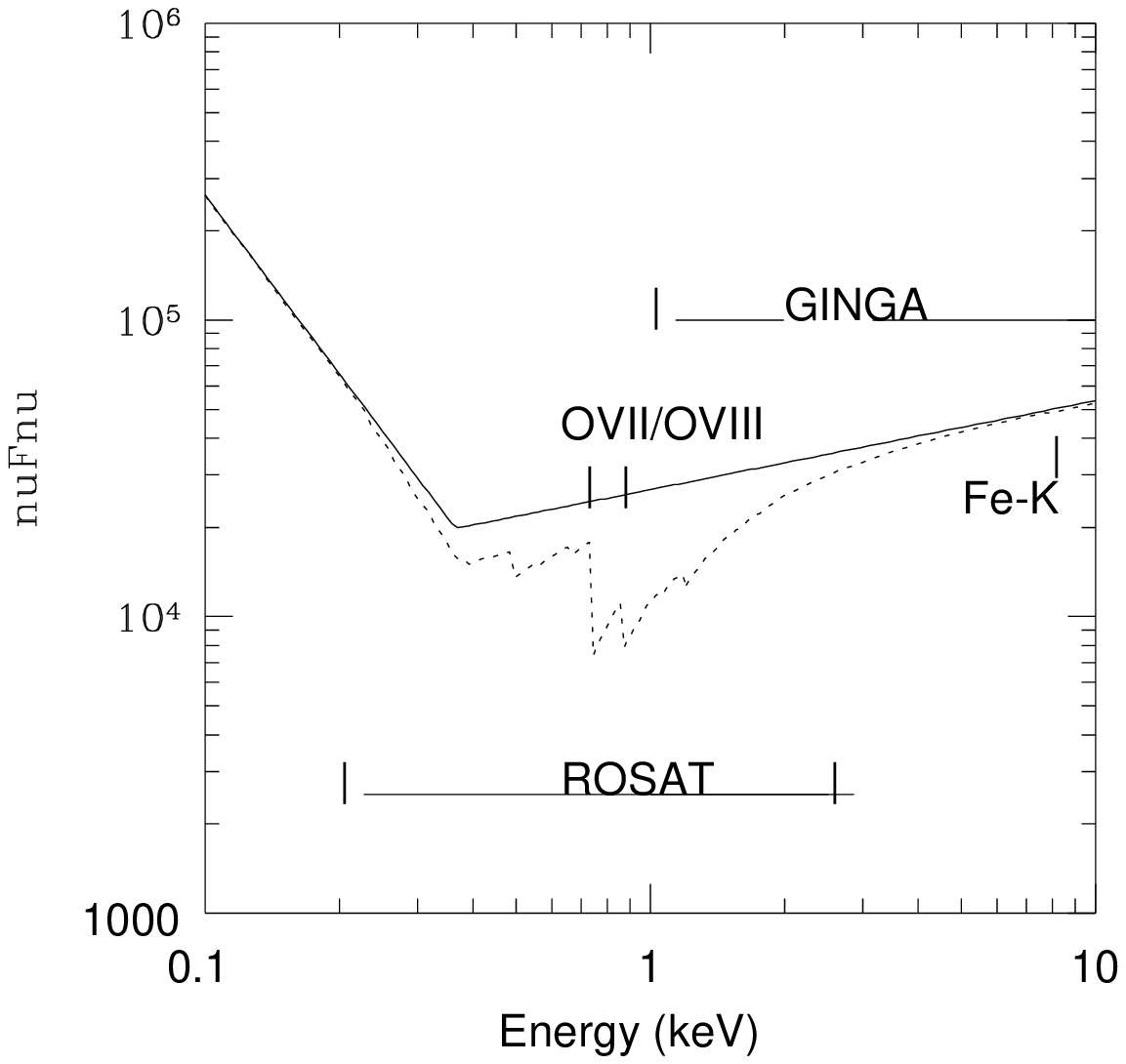}
}
\caption{  }
\end{figure}

\end{document}